\documentstyle[epsfig]{mn}
\begin{document}
\LARGE
\normalsize

\title[GRS 1915+105]
{Rapid infrared flares in GRS 1915+105 : evidence for infrared
synchrotron emission}

\author[R.~P.~Fender et al.]
{R. P. Fender$^1$\thanks{email : rpf@star.maps.susx.ac.uk}
G. G. Pooley$^2$, C. Brocksopp$^1$, 
S. J. Newell$^3$\\
$^1$ Astronomy Centre, University of Sussex, Falmer, Brighton BN1 9QH, UK\\
$^2$ Mullard Radio Astronomy Observatory, Cavendish Laboratory, 
Madingley Road, Cambridge CB3 OHE\\
$^3$ University of Manchester, Nuffield Radio Astronomy Laboratories,
Jodrell Bank, Macclesfield, Cheshire SK11 9DL, UK\\}

\maketitle

\begin{abstract}

We report imaging photometry of the radio-jet black hole candidate
source GRS 1915+105 in the infrared K band. The observations reveal
rapid infrared flare events on timescales of less than an hour. These
events are strikingly similar to those regularly observed in radio
monitoring at 15 GHz. Furthermore, when dereddened, the infrared
events have comparable amplitudes to the radio oscillations, and
observations at 15 GHz made $\sim 8$ hr after our infrared
observations reveal that the source was indeed displaying radio
oscillations at this time. We suggest that we have observed infrared
synchrotron emission from this source. We estimate the equipartition
magnetic field and power required to accelerate the particles
for the repeated radio events, and find both to be orders of
magnitude greater than those estimated for any other X-ray binary.

Comparison of events at 15 GHz and 2.2 $\mu$m suggests that the
dominant loss mechanism is adiabatic expansion, which in turn implies
that each event corresponds to a small ejection of material
from the system.

\end{abstract}

\begin{keywords}

binaries: close -- stars : individual : GRS 1915+105 -- infrared : stars
-- radio continuum : stars

\end{keywords}

\begin{figure}
\centering
\leavevmode\epsfig{file=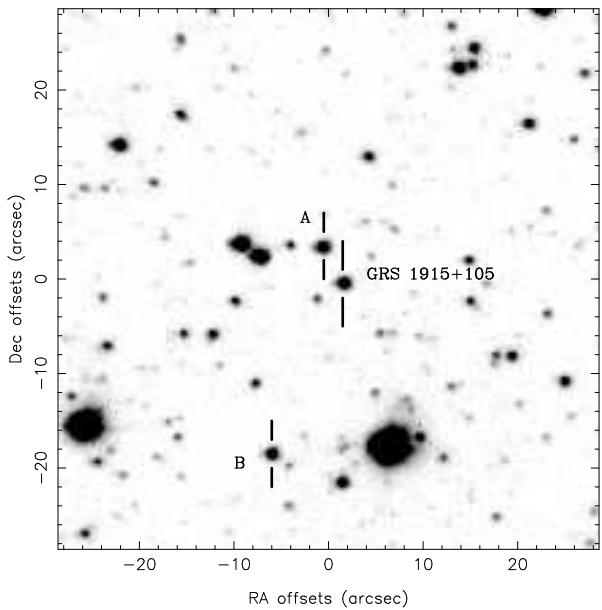,width=8cm,angle=270,clip}
\caption{Single 20-sec K-band image of GRS 1915+105 field, illustrating 
quality of images as well as locations of stars `A' and `B' which were
used for relative photometry.  There is no evidence for any extended
infrared structure associated with GRS 1915+105.}
\label{}
\end{figure}

\section{Introduction}

GRS 1915+105 was discovered by the WATCH instrument on board the
Granat satellite on 1992 August 15 (Castro-Tirado et al. 1992).  A
radio counterpart was subsequently identified, and in a series of
remarkable observations Mirabel \& Rodr\'{i}guez (1994) discovered
apparent superluminal motions in jet-like outflows from the source,
interpreted as a bipolar outflow with bulk velocity $\sim 0.9$c.
Radio monitoring of the source at 15 GHz has also revealed
quasi-periodic events with periods in the range 20 -- 120 min
(Pooley 1995, 1996, 1997, Pooley \& Fender 1997).

The source has no observed optical counterpart, due to heavy
extinction in the Galactic plane (see below), and its true nature
remains uncertain. The X-ray flaring and hard X-ray emission of the
source suggest a black-hole transient system (e.g. Finoguenov et
al. 1994), but the binarity of the system has yet to be definitively
established by any modulation at a likely orbital period.

GRS 1915+105 displays exceedingly rich behaviour in X-rays,
including very rapid large amplitude flaring on timescales
of minutes (e.g. Greiner, Morgan \& Remillard 1996; 
Belloni et al. 1997). The hard X-ray behaviour seems to
be coupled to the radio emission over long timescales
(Foster et al. 1996).

Spectroscopy of the infrared counterpart has revealed emission lines
of HeI and HI, and possibly also HeII, which have been interpreted as
arising in an accretion disc (Castro-Tirado, Geballe \& Lund 1996).
However, Mirabel et al. (1997) suggest that the emission lines may
arise in the circumstellar environment of a Be star in a high-mass
X-ray binary system.

\subsection{Previous infrared imaging and photometry}

The infrared counterpart to GRS 1915+105 was discovered independently
by Mirabel et al. (1994) and Castro-Tirado et al. (1993) in the J, H
\& K photometric bands. The K-magnitude was observed to vary between
13.0 -- 14.3, and the JHK colours suggested a heavily reddened
object. Bo\"{e}r et al. (1995) have detected a faint (I=23.4)
red-optical counterpart.

Sams, Eckart \& Sunyaev (1996a) made high-resolution K-band
observations with a speckle-imaging technique which revealed an
infrared `jet' of about 0.5 arcsec extent aligned at the same position
angle as the radio jet, when the source was in a bright (K=12.5)
state. Subsequent observations with the same technique, when the
source was again in a bright state (K=12.3) failed to reproduce this
result (Sams, Eckart \& Sunyaev 1996b). Eikenberry \& Fazio (1997) and
Chaty \& Mirabel (1997) describe further infrared imaging which shows no
evidence for a jet, and use this to place upper limits of the order of
days for the radiative lifetime of the emission.

Further photometry by Chaty et al. (1996) and Mirabel et al. (1996)
has revealed infrared variability in the J, H \& K bands on timescales
of hours and longer. In particular, Mirabel et al. (1996) report one
event in which GRS 1915+105 becomes brighter ($\Delta K \sim 1$ mag)
and redder ($\Delta (J-K) = 1.2$ mag) in the near-infrared a few days
after a major radio and X-ray outburst. They intepret this as
resulting from an interaction between jet ejecta and a dust cloud
lying several hundred A.U. from the source.

\section{Observations}

Our infrared observations were carried out at UKIRT on 1996 October
22, with IRCAM3, a $256\times256$ imaging array operating in the $1-5
\mu$m wavelength range. Conditions were photometric and seeing varying
between 0.8 -- 1.2 arcsec.  Over a $\sim 2.5$ hr period we obtained 145
K-band images of GRS 1915+105. Each image is a sum of five 4-sec
exposures, giving a total of 2900 sec on-source.  Effective time
resolution is of order 30 sec.  Flux calibration was achieved using
G21-15, for which we assumed K=11.757. The observations were carried
out in service mode, and were arranged to be simultaneous with pointed
OSSE hard X-ray observations (to be reported elsewhere).  Data
reduction was carried out using the {\small IRCAMDR} software on the
{\small STARLINK} node at Sussex University.

A single 20-sec K-band image of the field of GRS 1915+105 is shown in
Fig 1. The imaging is not of high enough resolution to look for the
infrared `jet' of Sams, Eckart \& Sunyaev (1996a). Indicated on the
image are GRS 1915+105 and two nearby stars of similar brightness,
labelled `A' and `B', which were used for relative photometry (see
below). Extended emission on arcsec scales, reported at radio
wavelengths by Newell, Spencer \& Jowett (1996), and our original
motivation for deep infrared imaging of the field, is not detected.

Photometry was performed on each of the 145 individual frames, with a
5 arcsec software aperture. Magnitudes were measured for GRS 1915+105,
star `A' and star `B' in every frame (except for a few where `B'
dropped off the edge of the image). Stars `A' and `B' then provided an
indication of how changing conditions were affecting the observations.
The mean magnitudes of `A' and `B' were $13.25 \pm 0.05$ and $13.34
\pm 0.05$ respectively.

Fig 2(a) shows the flux densities of GRS 1915+105, `A' and `B' plotted
against time. It is clear that `A' and `B' are changing together, as a
result of a varying atmosphere, and that the dramatic variability in
flux from GRS 1915+105 is real.  Dividing the flux from GRS 1915+105
by that of star `A' (Fig 2(b)) for each frame, and then multiplying by
the mean value for `A', provides a good compensation for varying
conditions. This is especially evident in the constant flux levels at
about 2.8 mJy (K=13.36) for GRS 1915+105 between flare events.

Radio observations reported here are drawn from a subset of those
to be presented in Pooley \& Fender (1997), in which the
observational procedure will be described in detail.

\begin{figure*}
\begin{minipage}{177mm}
\centering
\leavevmode\epsfig{file=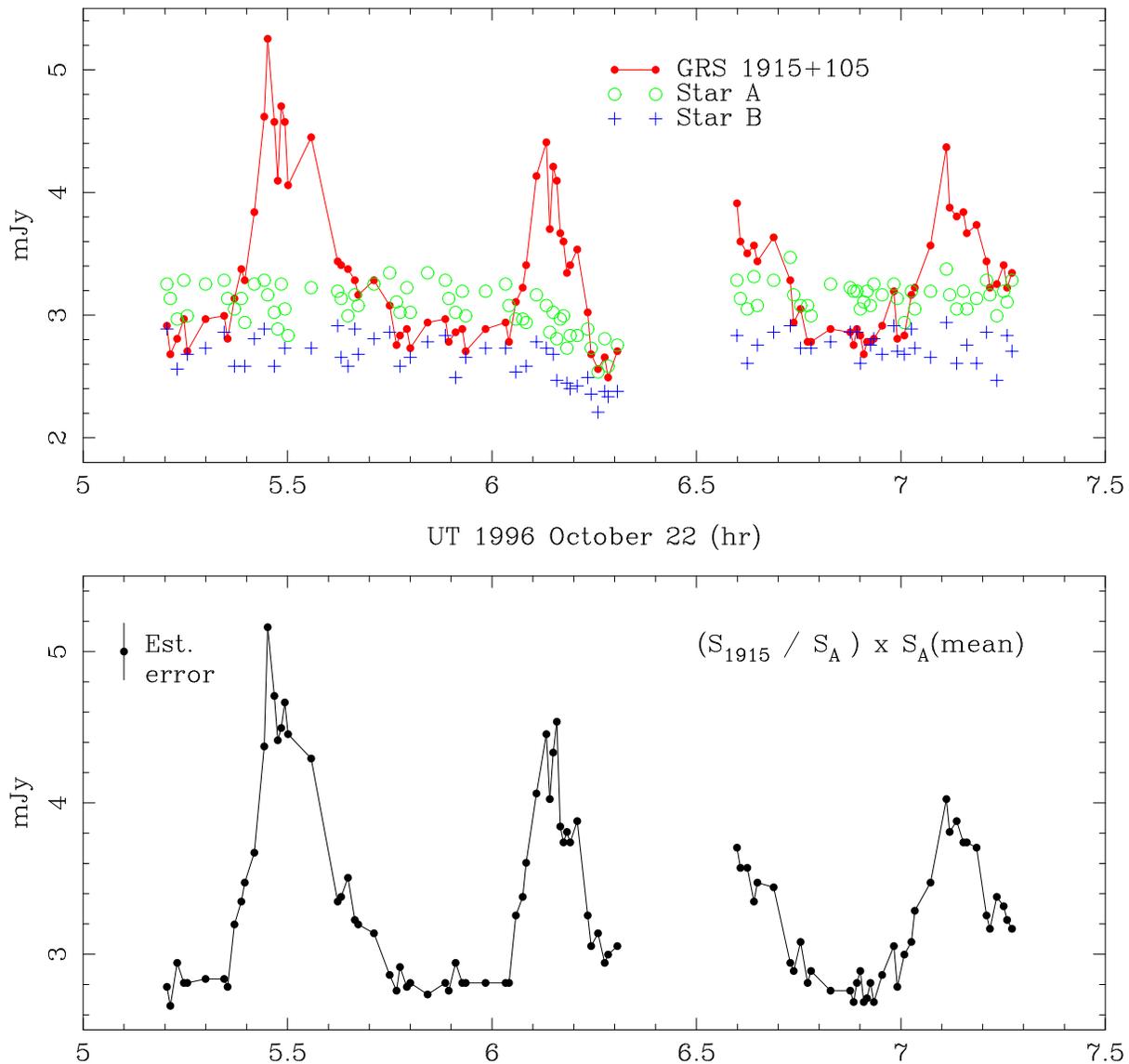,width=16cm}
\caption{Upper panel shows IRCAM3 K-band photometry of GRS 1915+105 and 
two nearby stars of similar brightness, labelled `A' and `B' in Fig
1. The measured flux densities of `A' and `B' clearly track each
other, indicating an origin in atmospheric fluctuations; the
variability in GRS 1915+105 is clearly real. The lower panel shows
the flux of GRS 1915+105 divided by that of star `A' for each frame, and
multiplied by mean flux density of star `A'. This successfully removes
atmospheric variability. There are clearly significant variations in 
the flux from GRS 1915+105 at the $\sim 30$ sec time resolution of
the observations, and a plateau level between flares of about 2.8 mJy.
The error estimate is the standard deviation in the measured flux
from Star A.}
\label{}
\end{minipage}
\end{figure*}

\begin{figure*}
\begin{minipage}{177mm}
\centering
\leavevmode\epsfig{file=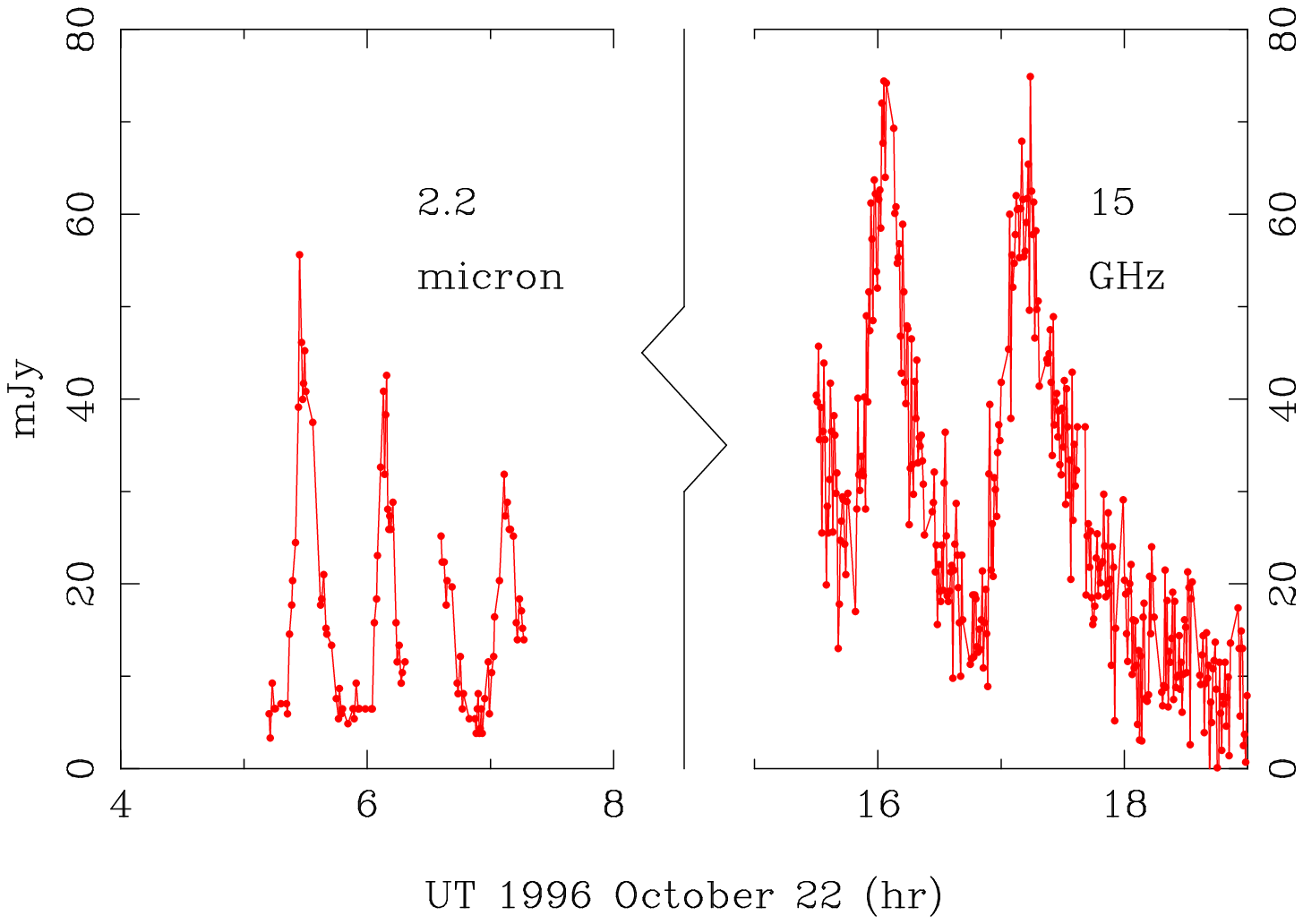,width=16cm}
\caption{Infrared flares from GRS 1915+105, dereddened by A$_{\rm K}$=3.3
and compared with radio observations which occurred 8 hr later.  A
baseline flux density of 2.5 mJy was subtracted from the infrared data
prior to dereddening.  The similarity in the behaviour at both
wavelengths is striking, suggesting a common emission process. A
brightness temperature in excess of 10$^{10}$ K for the emission at 15
GHz rules out a thermal origin; synchrotron emission seems the only
possible mechanism.  If synchrotron in origin, the energy required to
produce each infrared event is of order 10$^{40}$ erg.}
\label{}
\end{minipage}
\end{figure*}

\section{Discussion}

\subsection{Dereddening}

In order to compare the infrared flares meaningfully with both the
rapid radio variability repeatedly observed from GRS 1915+105, and
with similar infrared events observed in other sources (in particular
Cygnus X-3) it is necessary to deredden the observed flux
densities. Optical extinction to the source has been estimated to be
in the range 26 -- 33 magnitudes (Mirabel et al. 1994; Chaty et
al. 1997). This is consistent with K-band extinctions in the range 2.9
-- 3.7 mag (Rieke \& Lebofsky 1984) though the exact A$_V$:A$_K$
relation is uncertain and direction-dependent in the Galactic plane
(Mathis 1990).

We choose a value of A$_K$ = 3.3 mag (corresponding to A$_V \sim 30$
mag) to deredden the K-band fluxes by. However, it should be noted
that the uncertainty in the reddening to the source introduces
an uncertainty of at least 40\% in any quoted luminosities.

\subsection{Comparison with radio events}

Due to longitudinal differences, radio observations with the Ryle 
Telescope were not possible until $\sim 8$ hr after our infrared
observations. Figure 3 plots both the dereddened infrared flux
densities and the radio observations on the same scale. A value
of 2.5 mJy has been subtracted from the infrared data prior to
dereddening, as a quiescent baseline.

It is immediately obvious the radio oscillations of the type reported
by Pooley (1995, 1996, 1997) and Pooley \& Fender (1997) were occuring
at this time. The similarity between the variability in the infrared
and the radio is striking.  The {\em amplitude} of the events is very
similar, as are their {\em rise}, {\em decay} and {\em recurrence
timescales}. The first two infrared flares are separated by $\sim 40$
min, a `period' often found in the radio oscillations. The recurrence
timescale of the radio oscillations observed 8 hr later is actually
slightly longer, but `period' changes of this type are commonly
observed in the radio monitoring (Pooley \& Fender 1997).

\begin{table*}
\centering
\begin{minipage}{140mm}
\caption{Derived properties of the oscillation events 
under the assumption of a flat synchrotron spectrum from 15 GHz to 2.2 $\mu$m,
a volume limited by a rise time of 5 min and a filling factor of unity.}
\begin{tabular}{@{}cccc}
Synchrotron luminosity & Equipartition magnetic & Minimum energy in & Minimum power into \\
(erg s$^{-1}$)         & field (Gauss)          & each event (erg)  & particle acceleration (erg s$^{-1}$) \\
$\geq 10^{36}$ & $\sim 8$ & $10^{40}$ & $5 \times 10^{36}$ \\
\end{tabular}
\end{minipage}
\end{table*}

\subsection{Energy in synchrotron events}

If the infrared flares {\em are} synchrotron in origin, we can
evaluate the equipartition magnetic field for these events in order to
determine the minimum energy required for their generation. Assuming a
distance of 12.5 kpc and a flat spectrum (spectral index $\alpha = 0$)
from 15 GHz to 2.2 $\mu$m, the source has a synchrotron luminosity of
$\geq 10^{36}$ erg s$^{-1}$ during these outbursts (the luminosity is
completely dominated by the high-frequency infrared tail of the
emission). Assuming a volume for the electron cloud of a sphere of
radius 5 light min (from the rise time of the events), and a filling
factor of unity, we can estimate an equipartition magnetic field and
energy. The derived magnetic field is $\sim 8$ Gauss, the total energy
$\sim 10^{40}$ erg. While there is no reason to assume the source to
be in equipartition, shifting the magnetic field to higher or lower
values rapidly increases the energy associated with each oscillation;
we can therefore consider $10^{40}$ erg to be a lower limit on the
energy required to generate each event.

It should also be noted that our calculation is for an intrinsic
synchrotron spectral index of zero; the common interpretation of
a flat synchrotron spectrum as arising from an inhomogenous source
with varying degrees of absorption again increases the inferred
synchrotron luminosity and hence energy associated with each
oscillation. Furthermore, we do not consider effects naturally
associated with the jets, i.e. acceleration to relativistic
velocities and Doppler de-boosting, both of which will again
increase the required energy budget.

Assuming that the magnetic field does not decay, and can be re-used by
subsequent populations of accelerated particles, we find that GRS
1915+105 is required to input $\geq 5 \times 10^{36}$ erg s$^{-1}$
into the acceleration of particles in order to produce one such flare
event every $\sim$ 30 min (we stress that we have not considered any
bulk acceleration of material when arriving at this minimum power
requirement).  In a 8 G field, particles of energies $\sim 1$ GeV are
required to produce synchrotron emission at 2.2 $\mu$m. The lifetime
of such particles to synchrotron losses would be of order 1 hr,
significantly longer than that observed here.  Furthermore, the
characteristics timescale for flux density decays due to synchrotron
or inverse Compton losses are frquency dependent, varying as
$\nu^{0.5}$. If these were the dominant loss mechanisms, we would
therefore expect to see decay timescales $\sim 100$ times shorter at
2.2 $\mu$m than at 15 GHz. This is not however observed, and the
favoured decay mechanism must instead be adiabatic expansion losses,
which produce the same decay rate at all frequencies.

The observed and derived parameters of the oscillations are summarized
in table 1.

\subsection{Comparison with infrared flares in other objects}

The infrared flaring reported here is reminiscent of that observed in
Cygnus X-3 (e.g. Fender et al. 1996), in which flare complexes can
rise in less than a minute and persist for up to an hour. Two-colour
analysis of Cyg X-3 flares (Fender et al. 1996) suggested an origin in
a hot ($\geq 10^6$ K) optically thin plasma, possibly associated with
jets and/or the inner accretion disc (although infrared synchrotron
emission was not ruled out).  Given uncertainties in distance and
reddening to both sources, the infrared flares reported here are the
same luminosity as those observed in Cyg X-3.

Circinus X-1, related to GRS 1915+105 and Cyg X-3 in being a
radio-bright X-ray binary, undergoes infrared flaring once every 16.6
days (Glass 1994).  The flares are of longer duration (over a day)
than those observed in GRS 1915+105 and Cyg X-3, but again are
correlated with radio behaviour, and have no satisfactory explanation
to date.

\section{Conclusions}

We have observed GRS 1915+105 with high time resolution in the
infrared K-band and found flaring/oscillation events on timescales of
$\sim 30$ min which have striking morphological similarities to
oscillations often observed at high radio frequencies. When
dereddened, the infrared events are found to have (within
uncertainties primarily related to the dereddening) the same amplitude
as radio events observed from the source some eight hours later. We
suggest that the infrared and radio events have a common origin in
synchrotron emission. This supports the preferred explanation of Sams
et al. (1996a) that they were observing synchrotron emission from the
infrared jet, and is consistent with a reddening of the source during
outbursts without the need to include a dust component.

If the synchrotron interpretation is correct, then the synchrotron
luminosity of GRS 1915+105 becomes very large, of order 10$^{36}$ erg,
compared to other Galactic sources.  However, it should be noted that
previous estimates of the synchrotron luminosities of X-ray binaries
are based upon an extrapolation of optically thin spectra from cm
wavelengths, and the high-frequency form of the synchrotron spectrum
is not well known for these sources.  For GRS 1915+105, calculation of
equipartition magnetic field and minimum energy associated with an
oscillation, based upon the luminosity and size constraints implied by
the rise times, produces values of $\sim 8$ G and $\sim 10^{40}$ erg
respectively.

The dominant mechanism causing the decay of the emission is determined
to be adiabatic expansion losses from the similarity of characteristic
decay times at two wavelengths four orders of magnitude apart. This in
turn implies that each oscillation event corresponds to a small
ejection of material from the system, and is not consistent with
radiative losses suffered by a confined emission region. Synchrotron
losses will become more important for observations at shorter
wavelengths. For the equipartition magnetic field of $\sim 8$ G, the
radiative lifetime would be shorter than the expansion lifetime at
wavelengths of $\sim 0.2$ $\mu$m and shorter, producing a turnover in
the synchrotron spectrum during the decay phase of any oscillations.

\section*{Acknowledgements}

We would like to thank Vivek Dhawan, Bob Hjellming and Rob Hynes for
useful discussions. We would like to thank the service team at UKIRT,
which is operated by The Observatories on behalf of the PPARC, and the
staff at MRAO for maintenance and operation of the Ryle Telescope,
which is supported by the PPARC.

\end{document}